\documentclass[aps,twocolumn,prl,floatfix,superscriptaddress,showpacs]{revtex4-1}
\usepackage{amssymb}

\usepackage{epsf}
\usepackage{epsfig}
\usepackage{graphicx}
\usepackage{dcolumn}
\usepackage{braket}
\usepackage{bm}
\usepackage{amsfonts}
\usepackage{amsmath}
\usepackage{amssymb}
\usepackage{color,soul}
\usepackage{wasysym}
\usepackage{mathrsfs}
\usepackage{times}
\usepackage{hyperref,fancyhdr,enumitem,color}

\newcommand{\bea}{\begin{eqnarray}}
\newcommand{\eea}{\end{eqnarray}}

\newcommand{\ci}{\mathrm{i}}

\newcommand{\overbar}[1]{\mkern 1.5mu\overline{\mkern-1.5mu#1\mkern-1.5mu}\mkern 1.5mu}

\begin{document}

\title{Multiterminal Conductance of a Floquet Topological Insulator}
\author{L. E. F. Foa Torres}%$^*$
%\email{lfoa@famaf.unc.edu.ar}
\affiliation{Instituto de F\'{\i}sica Enrique Gaviola (CONICET) and FaMAF, Universidad Nacional de C\'ordoba, Argentina}
\affiliation{Consejo Nacional de Investigaciones Cient\'{\i}ficas y T\'ecnicas (CONICET), Argentina}
\author{P. M. Perez-Piskunow}
\affiliation{Instituto de F\'{\i}sica Enrique Gaviola (CONICET) and FaMAF, Universidad Nacional de C\'ordoba, Argentina}
\affiliation{Consejo Nacional de Investigaciones Cient\'{\i}ficas y T\'ecnicas (CONICET), Argentina}
\author{C. A. Balseiro}
\affiliation{Centro At{\'{o}}mico Bariloche and Instituto Balseiro,
Comisi\'on Nacional de Energ\'{\i}a At\'omica, 8400 Bariloche, Argentina}
\affiliation{Consejo Nacional de Investigaciones Cient\'{\i}ficas y T\'ecnicas (CONICET), Argentina}
\author{G. Usaj}
\affiliation{Centro At{\'{o}}mico Bariloche and Instituto Balseiro,
Comisi\'on Nacional de Energ\'{\i}a At\'omica, 8400 Bariloche, Argentina}
\affiliation{Consejo Nacional de Investigaciones Cient\'{\i}ficas y T\'ecnicas (CONICET), Argentina}
%\email{usaj@cab.cnea.gov.ar}

\begin{abstract}
We report on simulations of the dc conductance and quantum Hall response of a Floquet topological insulator using Floquet scattering theory. Our results reveal that laser-induced edge states in graphene lead to quantum Hall plateaus once imperfect matching with the non-illuminated leads is lessened. But the magnitude of the Hall plateaus is not directly related to the number and chirality of all the edge states at a given energy as usual. Instead, the plateaus are dominated only by those edge states adding to the dc density of states. Therefore, the dc quantum Hall conductance of a Floquet topological insulator is not directly linked to topological invariants of the full the Floquet bands.

\end{abstract}
% 73.22.Pr   	Electronic structure of graphene 
% 73.20.At 	Surface states, band structure, electron density of states 
% 72.80.Vp	Electronic transport in graphene 
% 73.63.-b      Electronic transport, nanoscale materials
% 78.67.-n   	Optical properties of low-dimensional, mesoscopic, and nanoscale materials and structures
% 73.43.-f	quantum Hall effects,
% 73.43.Nq	Phase transitions quantum Hall effects
% 05.60.Gg	Transport processes, quantum
\date{28 August 2014}
\maketitle

\textit{Introduction.--} Floquet topological insulators (FTIs) \cite{Oka2009,Kitagawa2010,Lindner2011} are an incarnation of topological insulators (TIs) \cite{Kane2005,Koenig2007,Hasan2010} where the non-trivial topological properties \cite{Oka2009} are crafted with an external driving (\textit{e.g.} a circularly polarized laser). In FTIs, the Floquet chiral edge states bridge either a native bulk gap, as in semiconductor quantum wells \cite{Lindner2011}, or a gap which is also produced by the driving, as in the case of graphene \cite{Perez-Piskunow2014,Usaj2014}. Recently, laser-induced gaps have been probed at the surface of a 3d topological insulator \cite{Wang2013}. The field is evolving at a fast pace \cite{Gu2011,Gomez-Leon2013,Kundu2013,Thakurathi2013} with additional facets in general quantum physics \cite{Rudner2013,Ho2014,Asboth2014}, cold atoms \cite{Atala2013,Goldman2014,Choudhury2014,Jotzu2014} and photonic crystals \cite{Rechtsman2013}.

The search for Floquet topological states has started and some theoretical proposals \cite{Oka2009,Kitagawa2011,SuarezMorell2012,Perez-Piskunow2014,Sentef2014} embrace a realization in broadly available materials such as graphene \cite{Novoselov2005a,CastroNeto2009}. These states could be probed through pump-probe photoemission \cite{Sentef2014}, or STM \cite{Fregoso2014b}. But a crucial issue remains: the connection between the Floquet quasi-energy spectra and the conductance.

Indeed, one of the theoretical milestones established shortly after the discovery of the quantized Hall effect \cite{vonKlitzing1980} is the connection between the Hall conductance and a topological invariant \cite{Thouless1982} (the Chern numbers). This invariant, in turn, is related to the chiral edge states through the bulk-boundary correspondence \cite{Hasan2010,Bernevig2013}. For FTIs, the situation is more subtle. On one hand the non-equilibrium electronic occupations \cite{Oka2009,Dehghani2014} pose a difficult problem if dissipation is to be considered within the system \cite{Hone2009,Dehghani2014}. An alternative is a setup where external driving is limited to a finite region, thus leaving well defined occupations for the asymptotic states \cite{Moskalets2002,Camalet2003,Kohler2005},  
which can be handled through a scattering approach \cite{Gu2011,Kitagawa2011}. But even in this case, while some authors argue that the Hall conductance will be quantized within a few percents of $2e^2/h$ \cite{Kitagawa2011}, others claim that the two-terminal dc conductance may show an anomalous suppression \cite{Kundu2014}. On the other hand, there could also be a dc current at zero bias voltage (a pumping current), thereby complicating the expected transport response.

\begin{figure}[tb]
\includegraphics[width=\columnwidth]{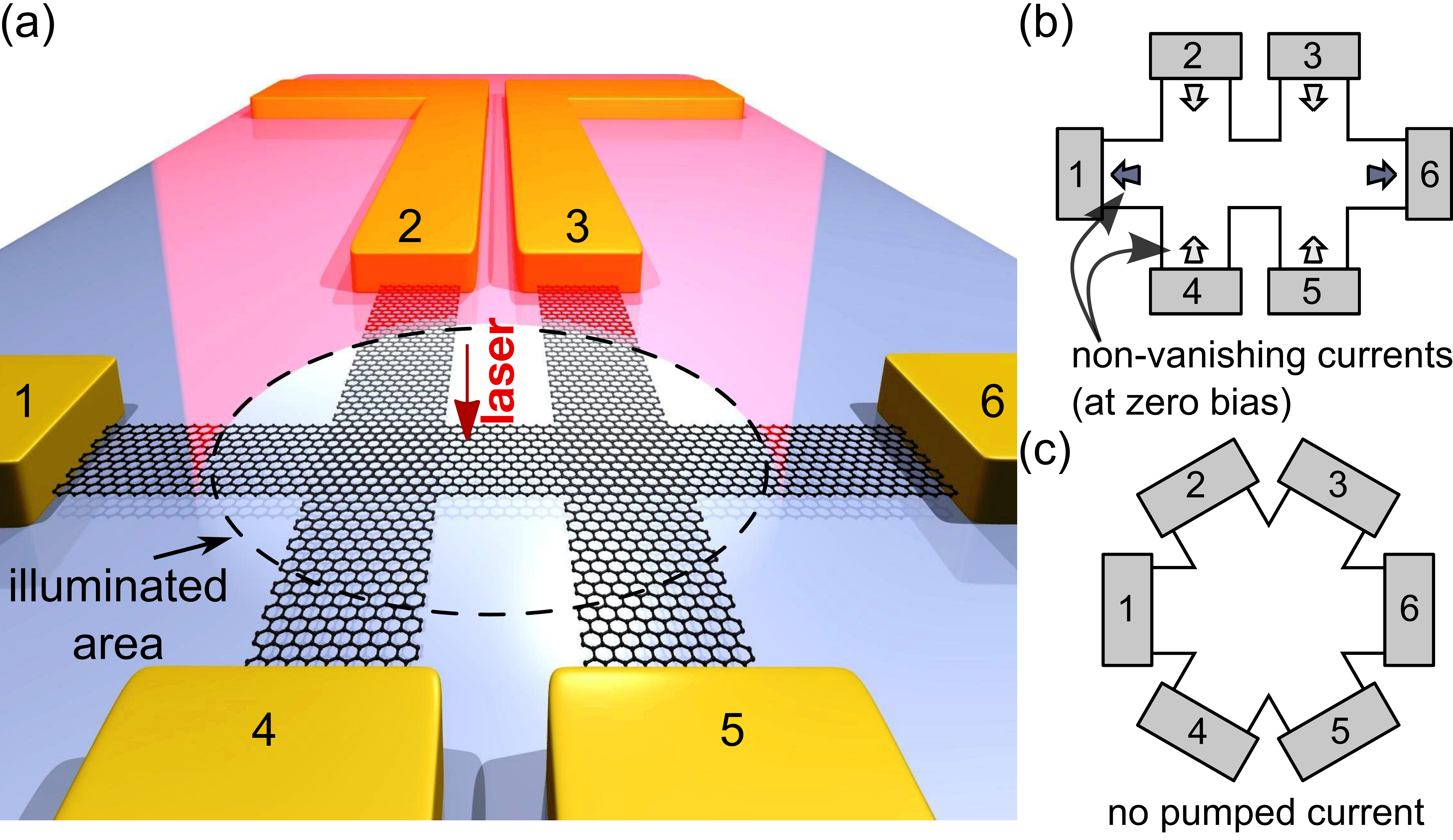}
\caption{
(a) Scheme of a setup where a laser of frequency $\Omega$ illuminates part of a sample in a six-terminal configuration. (b) Six-terminal setup with arrows representing one of the possible directions for the currents induced at zero bias voltage (pumped currents) on each lead.  
(c) Six-terminal configuration in an hexagonal arrangement. For a graphene sample, this high symmetry configuration gives a vanishing pumped current. }
\label{fig1}
\end{figure}

Here, we explicitly address dc charge transport in FTIs in a multi-terminal geometry (Fig. \ref{fig1}). We use graphene as an example for explicit calculations but our main conclusions are general. First, by using a Floquet scattering picture \cite{Moskalets2002,Camalet2003,Kohler2005} we discuss the calculation of the dc conductance and address the role of the voltmeters in driven systems where a current may arise even at zero bias. Later on we turn to the simulations for illuminated graphene. Our results show that the \textit{non-local} dc Hall conductance can reach roughly constant values, once the setup is tuned to lessen the imperfect matching between the irradiated area and the non-irradiated leads. More important, we find a major departure from other topological systems: the magnitude of the Hall conductance plateaus are determined only by a subset of all the Floquet chiral edge states available at a given energy. This breaks down the connection between topological invariants such as the winding numbers \cite{Rudner2013} and the Hall plateaus in FTIs.

\textit{Floquet theory and dc current.--} Floquet theory offers a suitable framework for systems driven by a time-periodic perturbation \cite{Grifoni1998,Kohler2005}. One starts by noting that for a Hamiltonian $\hat{\cal{H}}$ with a time period $T$, there is a complete set of solutions of the form $\psi_{\alpha}(\bm{r},t)=\exp(-\ci\varepsilon_{\alpha}t/\hbar) \phi_{\alpha}(\bm{r},t)$, where $\varepsilon_{\alpha}$ are the so-called quasienergies and $\phi_{\alpha}(\bm{r},t+T)=\phi_{\alpha}(\bm{r},t)$ are the Floquet states. By replacing these solutions into the time-dependent Schr\"odinger equation the Floquet states turn out to satisfy an equation analogous to the time-independent Schr\"odinger equation with the Hamiltonian being replaced by the Floquet Hamiltonian $\hat{\cal{H}}_F\equiv\hat{\cal{H}}-\ci\hbar \frac{\partial}{\partial t}$. Therefore, one has an eigenvalue problem in the direct product (Floquet) space \cite{Sambe1973}: $\mathscr{R}\otimes \mathscr{T}$, $\mathscr{R}$ being the usual Hilbert space and $\mathscr{T}$ the space of periodic functions with period $T=2\pi/\Omega$.

When the time-dependent potential is limited to the \textit{scattering region} (either because of a finite laser spot or the screening inside metallic contacts) as in Fig. \ref{fig1}(a) , the asymptotic states and their occupations remain well defined. One has a coherent scattering picture \cite{Moskalets2002,Camalet2003} where dissipation is assumed to take place far in the leads. In a multi-terminal setup, the time-averaged current at lead $\alpha$, $\overbar{{\cal I}}_{\alpha}=\frac{1}{T}\int_0^T {\cal I}_{\alpha}(t) dt$, is \cite{Moskalets2002}:
%%%%%%%%%%%%%%%%%%%%%%%%%%%%%%%%%%%%%%%%%%%%%%%%%
%\begin{eqnarray}
%\label{Floquet-Current}
%\overbar{{\cal I}}_{\alpha}&=&\frac{1}{T}\int_0^T {\cal I}_{\alpha}(t) dt\\ 
%\nonumber
%&=&\frac{2e}{h}\sum_{\beta\neq\alpha}\sum_{n}\!\int\!\left[  {\cal T}_{\beta,\alpha}^{(n)}%
%(\varepsilon)f_{\alpha}(\varepsilon)-{\cal T}_{\alpha,\beta}^{(n)}(\varepsilon
%)f_{\beta}(\varepsilon)\right]  d\varepsilon.
%\end{eqnarray}
\begin{equation}
\label{Floquet-Current}
\overbar{{\cal I}}_{\alpha}\!=\!\frac{2e}{h}\sum_{\beta\neq\alpha}\sum_{n}\!\int\!\left[  {\cal T}_{\beta,\alpha}^{(n)}(\varepsilon)f_{\alpha}(\varepsilon)\!-\!{\cal T}_{\alpha,\beta}^{(n)}(\varepsilon)f_{\beta}(\varepsilon)\right]  d\varepsilon\,.
\end{equation}
%%%%%%%%%%%%%%%%%%%%%%%%%%%%%%%%%%%%%%%%%%%%%%%%%
${\cal T}_{\beta,\alpha}^{(n)}(\varepsilon)$ is the transmission probability for an electron from lead $\alpha$ with energy $\varepsilon$  to lead $\beta$ emitting (absorbing) $n>0$ ($n<0$) photons and $f_{\alpha}(\varepsilon)$ is the Fermi functions at lead $\alpha$. In the absence of many-body interactions this is equivalent to the Keldysh formalism \cite{Kohler2005,Arrachea2006}. 

The differential conductance can be obtained after a linear expansion in the bias voltage(s). But there is a key issue to bear in mind: The time-dependent potential may break reciprocity, \textit{i.e.} ${\cal T}_{\alpha,\beta} \neq {\cal T}_{\beta,\alpha}$ (${\cal T}_{\alpha,\beta}\equiv \sum_{n}{\cal T}_{\alpha,\beta}^{(n)}$ is the total transmission probability). This is what happens, for example, when a magnetic field is present. However, here this inequality also holds for the sum, $\sum_{\beta}{\cal T}_{\beta,\alpha}(\varepsilon) \neq \sum_{\beta}{\cal T}_{\alpha,\beta}(\varepsilon)$,
%$\sum_{\beta\neq\alpha}{\cal T}_{\beta,\alpha}(\varepsilon) \neq \sum_{\beta\neq\alpha}{\cal T}_{\alpha,\beta}(\varepsilon)$,
%%%%%%%%%%%%%%%%%%%%%%%%%%%%%%%%%%%%%%%%%%%%%%%%%
%\begin{equation}
%\sum_{\beta\neq\alpha}{\cal T}_{\beta,\alpha}(\varepsilon) \neq \sum_{\beta\neq\alpha}{\cal T}_{\alpha,\beta}(\varepsilon)\,,
%\label{Sum-rule-zero-current-at-eq}
%\end{equation}
%%%%%%%%%%%%%%%%%%%%%%%%%%%%%%%%%%%%%%%%%%%%%%%%%
\textit{i.e.} the current is generally non-zero even in the absence of a bias-voltage. 
This `pumping' contribution \cite{Moskalets2002} is missing in the usual linearization and will be considered below.

\begin{figure}[tbp]
\includegraphics[width=0.9\columnwidth]{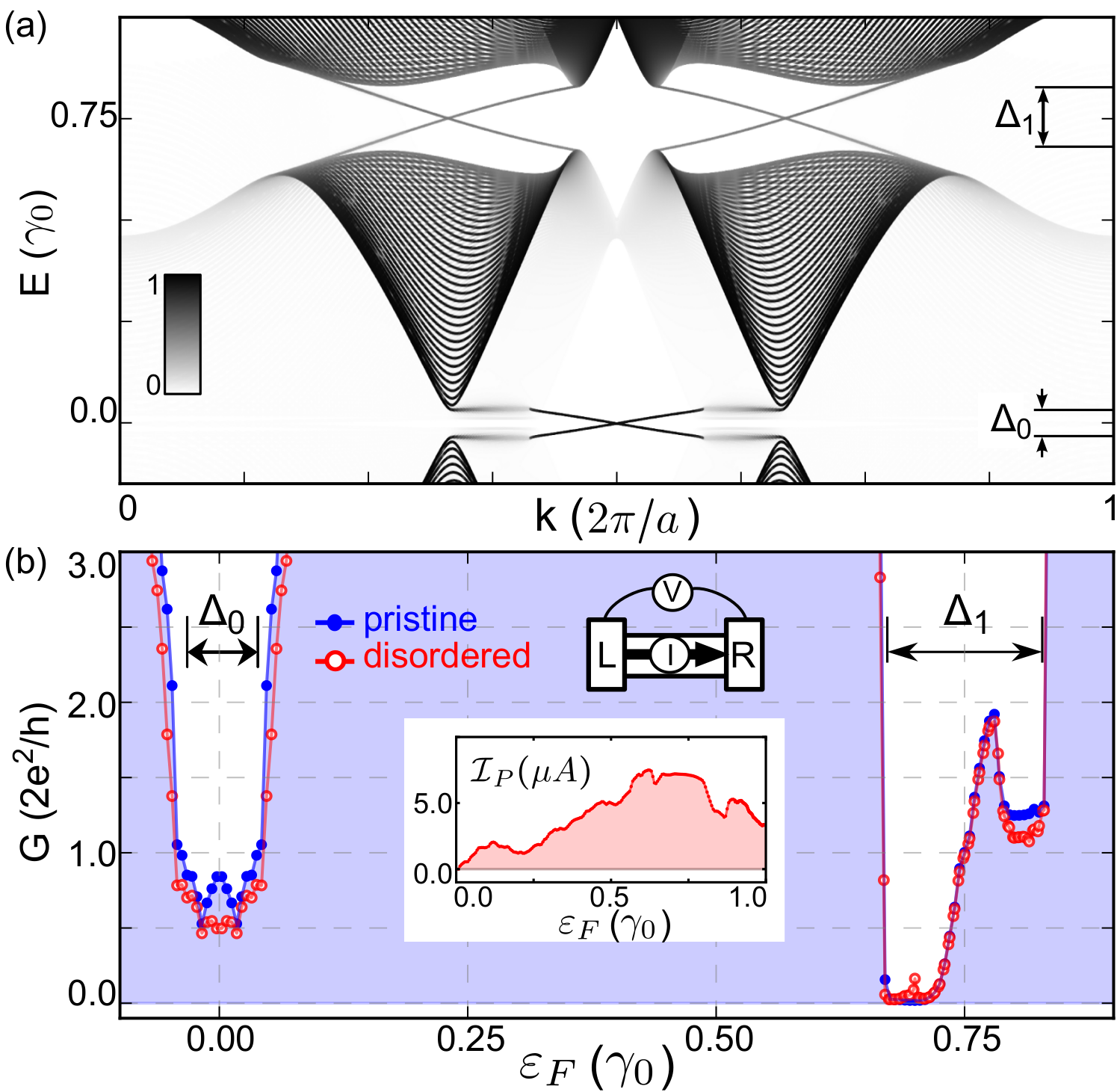}
\caption{
(a) (color online) Quasi-energy dispersion for a zigzag graphene ribbon of width $W=125a$ illuminated by a circularly polarized laser of frequency $\hbar\Omega=1.5\gamma_0$ and intensity $z=0.15$. The color scale encodes the weight of the state on the dc density of states: white for no weight and black maximum weight. The numerics consider Floquet replicas with $n$ between $-2$ and $2$. (b) Conductance in a two-terminal setup where a section of the infinite zigzag ribbon of length $L=420a$ is illuminated. Full (blue) circles correspond to the pristine system while the empty (red) circles are for a ribbon with 15 vacancies distributed at random in the illuminated area. Symmetry breaking due to defects leads to a pumped current $\mathcal{I}_P$ (inset).}
\label{fig2}
\end{figure}

\textit{Conductance in a two-terminal setup.--} In this case $\alpha,\beta=L,R$ (left, right) and unitarity requires $\overbar{\cal{I}}_L=-\overbar{\cal{I}}_R\equiv\overbar{\cal{I}}$. Eq. (\ref{Floquet-Current}) can be written as
%%%%%%%%%%%%%%%%%%%%%%%%%%%%%%%%%%%%%%%%%%%%%%%%%
%\begin{eqnarray}
%\overbar{{\cal I}}=\frac{2e}{h}\int\Big[&&{\cal T}(\varepsilon) \big( f_{L}(\varepsilon)-f_{R}(\varepsilon)\big)\\
%\nonumber 
%&&+\delta {\cal T}^{}(\varepsilon) \big(f_{L}(\varepsilon)+f_{R}(\varepsilon)\big)\Big]  d\varepsilon,
%\label{Floquet-Current-2-terminals}
%\end{eqnarray}
\begin{equation}
\overbar{{\cal I}}=\frac{2e}{h}\!\int\!\Big[{\cal T}(\varepsilon) [f_{L}(\varepsilon)-f_{R}(\varepsilon)]+\delta {\cal T}^{}(\varepsilon) [f_{L}(\varepsilon)+f_{R}(\varepsilon)]\Big]  d\varepsilon,
\label{Floquet-Current-2-terminals}
\end{equation}
%%%%%%%%%%%%%%%%%%%%%%%%%%%%%%%%%%%%%%%%%%%%%%%%%
where $\mathcal{T}(\varepsilon)=[{\cal T}_{R,L}(\varepsilon) + {\cal T}_{L,R}(\varepsilon)]/2$ and $\delta {\cal T} = ({\cal T}_{R,L} - {\cal T}_{L,R})/2$. %characterizes the asymmetry of the transmission coefficients. 
%For simplicity 
We consider the zero temperature limit but generalization to finite temperature is direct. 
To linear order in the bias voltage $V$ we get $\overbar{{\cal I}}\simeq \frac{2e^2}{h} {\cal T}(\varepsilon_F) \times V + \frac{e^2}{h} \int \delta {\cal T}(\varepsilon)f(\varepsilon)  d\varepsilon$.
%%%%%%%%%%%%%%%%%%%%%%%%%%%%%%%%%%%%%%%%%%%%%%%%%
%\begin{equation}
%\overbar{{\cal I}}\simeq \frac{2e^2}{h} {\cal T}(\varepsilon_F) \times V + \frac{e^2}{h} \int \delta {\cal T}(\varepsilon)f(\varepsilon)  d\varepsilon\, .
%\label{Floquet-Current-2-terminals-linear}
%\end{equation}
%%%%%%%%%%%%%%%%%%%%%%%%%%%%%%%%%%%%%%%%%%%%%%%%%
The second term can be interpreted as a \textit{pumping current} ($\mathcal{I}_P$) resulting from the asymmetry of the transmission coefficients. Since it does not depend on the bias voltage, the differential conductance is $(2e^2/h){\cal T}(\varepsilon_F)$. In contrast to the result for time-independent systems, the conductance depends on the transmission probabilities from left to right \textit{and} from right to left. 
When inversion symmetry holds, both leads are indistinguishable ($\delta {\cal T}=0$) and unitarity warrants zero current at zero bias. Breaking the symmetry either because of defects or slightly different contacts may introduce a large asymmetry in the transmission coefficients as we will see later on when discussing Fig. \ref{fig2}.

\textit{Multi-terminal conductance.--} 
In contrast to the two-terminal case, in a multi-terminal setup inversion symmetry (IS) does not warrant a zero pumped current (contrary to what was argued in Ref. [\onlinecite{Kitagawa2011}]). Indeed, in a six terminals configuration as the one in Fig. \ref{fig1}(b), in the absence of any bias voltage inversion symmetry requires $\overbar{{\cal I}}_1=\overbar{{\cal I}}_6$ and $\overbar{{\cal I}}_2=\overbar{{\cal I}}_5$ and $\overbar{{\cal I}}_3=\overbar{{\cal I}}_4$ which together with charge conservation may lead to solutions where, for example, leads $2$ to $5$ feed non-zero currents into leads $1$ and $6$. Therefore, unless more stringent conditions on the geometry of the system are imposed, the voltmeter may have to do additional work against the time-dependent field to keep a zero dc current.

To measure a multi-terminal differential conductance we need to establish a protocol. One starts with no bias voltage and define a subset of electrodes as our voltmeters \cite{Buettiker1986,DAmato1990,Datta1995}. In our discussion we assume that a voltmeter is a device which adjusts its internal parameters to keep a zero dc current on the corresponding lead. 
In any case, the internal voltmeter parameters are shifted until the associated currents (given by Eq. (\ref{Floquet-Current})) are zero \cite{nota1}. Then one has a non-equilibrium state where we have counteracted the pumping currents at the voltmeters (the source and drain may still sustain a pumping current $\mathcal{I}_{P\alpha}$). This state is characterized by a set of internal parameters (that we omit for brevity) and chemical potentials on each lead denoted by $\{\mu_{\alpha}^{(0)}\}$. Starting from this state and assuming that deviations $\delta\mu_{\alpha}\equiv\mu_{\alpha}-\mu_{\alpha}^{(0)}$ are small enough one gets:
%%%%%%%%%%%%%%%%%%%%%%%%%%%%%%%%%%%%%%%%%%%%%%%%%
\begin{equation}
\overbar{\cal{I}}_{\alpha}=\mathcal{I}_{P\alpha}+ \frac{2e}{h}\sum_{\beta\neq\alpha}\big[  {\cal T}_{\beta,\alpha}(\varepsilon_{F})\delta\mu_{\alpha}%
-{\cal T}_{\alpha,\beta}(\varepsilon_{F})\delta\mu_{\beta}\big]\,,
\label{Floquet-multiterminal-linear-V}
\end{equation}
%%%%%%%%%%%%%%%%%%%%%%%%%%%%%%%%%%%%%%%%%%%%%%%%%
where $\mathcal{I}_{P\alpha}=0$ at the voltmeters. Then we proceed as usual: fixing a small bias between two leads and imposing a zero-dc current on the voltmeters. The obtained $\{\delta\mu_{\alpha}\}$ determine the conductance and Hall resistance.
Hence, though it may give a remnant current at zero bias, pumping should not affect the linear conductance. When \textit{all leads are equivalent} (a condition depending also on the details of the lattice), as in the setup of Fig. \ref{fig1}(c), the pumped currents vanish. 

\textit{Conductance of irradiated graphene ribbons.--} Illuminating graphene with circularly polarized light can turn it into a Floquet Topological Insulator: it develops bulk bandgaps \cite{Oka2009,Calvo2011} (both at the Dirac point and at $\pm \hbar\Omega/2$) which is bridged by chiral edge states \cite{Perez-Piskunow2014,Usaj2014}. The two-terminal conductance of these states has not been studied in detail, except at the Dirac point \cite{Gu2011,Dehghani2014}. 
But, to the best of our knowledge, apart from a calculation based on a Kubo formula presented in Ref. [\onlinecite{Oka2009}] and an approximate analytical calculation in Ref. [\onlinecite{Kitagawa2011}] no explicit results in a multi-terminal configuration for the \textit{non-local} conductance are available. We address this in the following.

We consider an all-graphene system where semi-infinite graphene ribbons serve as electrodes. Graphene is modeled through a standard $\pi$-orbitals Hamiltonian: ${\cal H}=\sum_{i}E_{i}^{{}}\,c_{i}^{\dagger}c_{i}^{{}}-\sum_{\left\langle i,j\right\rangle }[\gamma_{ij}c_{i}^{\dagger}c_{j}^{{}}+\mathrm{h.c.}]$, where $c_{i}^{\dagger}$ and $c_{i}^{{}}$ are the electronic creation and annihilation operators at the $\pi$-orbital localized on site $i$ which has energy $E_i$. $E_i$ are set equal to zero for the undoped system, doping can be included through a shift in $E_i$. $\gamma_{ij}=\gamma_0=2.7$ eV is the nearest-neighbors hopping \cite{FoaTorres2014}. 
The interaction with the laser (assuming normal incidence) is introduced through a time-dependent phase in the hopping amplitudes  \cite{Oka2009,Calvo2011},
$\gamma_{ij}=\gamma_{0}\exp\left(\mathrm{i}\frac{2\pi}{\Phi_0}\int_{\bm{r}_j}^{\bm{r}_i}\bm{A}(t)\cdot\mathrm{d}\bm{\ell}\right)$, where $\Phi_0$ is the magnetic flux quantum, $\bm{A}$ is the vector potential which is related to the electric field $\bm{E}$ through $\bm{E}=- (1/c)\partial \bm{A}/ \partial t$. The driving is assumed to take place only in the central region and is smoothly turned-off before reaching the leads. $z \equiv 2\pi a A_0/(\sqrt{3}\Phi_0)$ characterizes the laser strength, $A_0=|\bm{A}|$ and $a$ is the graphene lattice constant.
For the calculation of the probabilities and other magnitudes we used both an implementation built on Kwant \cite{Groth2014} and Floquet-Green's functions \cite{Calvo2011,FoaTorres2014}. Simulations for realistic values of the laser intensities/frequencies require large systems \cite{Perez-Piskunow2014,Usaj2014} with a high computing cost. Instead, the numerics here are aimed at illustrating fundamental issues of FTIs.

Figure \ref{fig2}(a) shows the quasi-energy dispersion of a zigzag ribbon illuminated by a circularly polarized laser. The laser opens bulk gaps both at the Dirac point and at $\pm \hbar\Omega/2$ but also produces chiral edge states bridging them.
Fig. \ref{fig2}(b) shows the dc two-terminal conductance for both the pristine system (full (blue) circles with a dashed area underneath) and the same ribbon with $15$ random vacancies within the illuminated area (empty (red) circles). As expected, there is a strong reduction of the conductance at the laser induced-gaps: the conductance within these bulk gaps is due to the chiral edge states. While the magnitude of the latter has the correct order of magnitude ($\sim 2e^2/h$),  it presents a strong modulation as a function of energy. We will come back to this later. In addition, the presence of disorder gives a directional asymmetry: 
in spite of the small number of vacancies (less than $1\permil$), there is an asymmetry in the transmission coefficients, 
resulting in a pumped current (Fig. \ref{fig2}(b) inset). We find that this effect is even stronger in the H-shaped setup of Fig. \ref{fig1}(b) even without disorder \cite{supp-info}.

\begin{figure}[t]
\includegraphics[width=0.9\columnwidth]{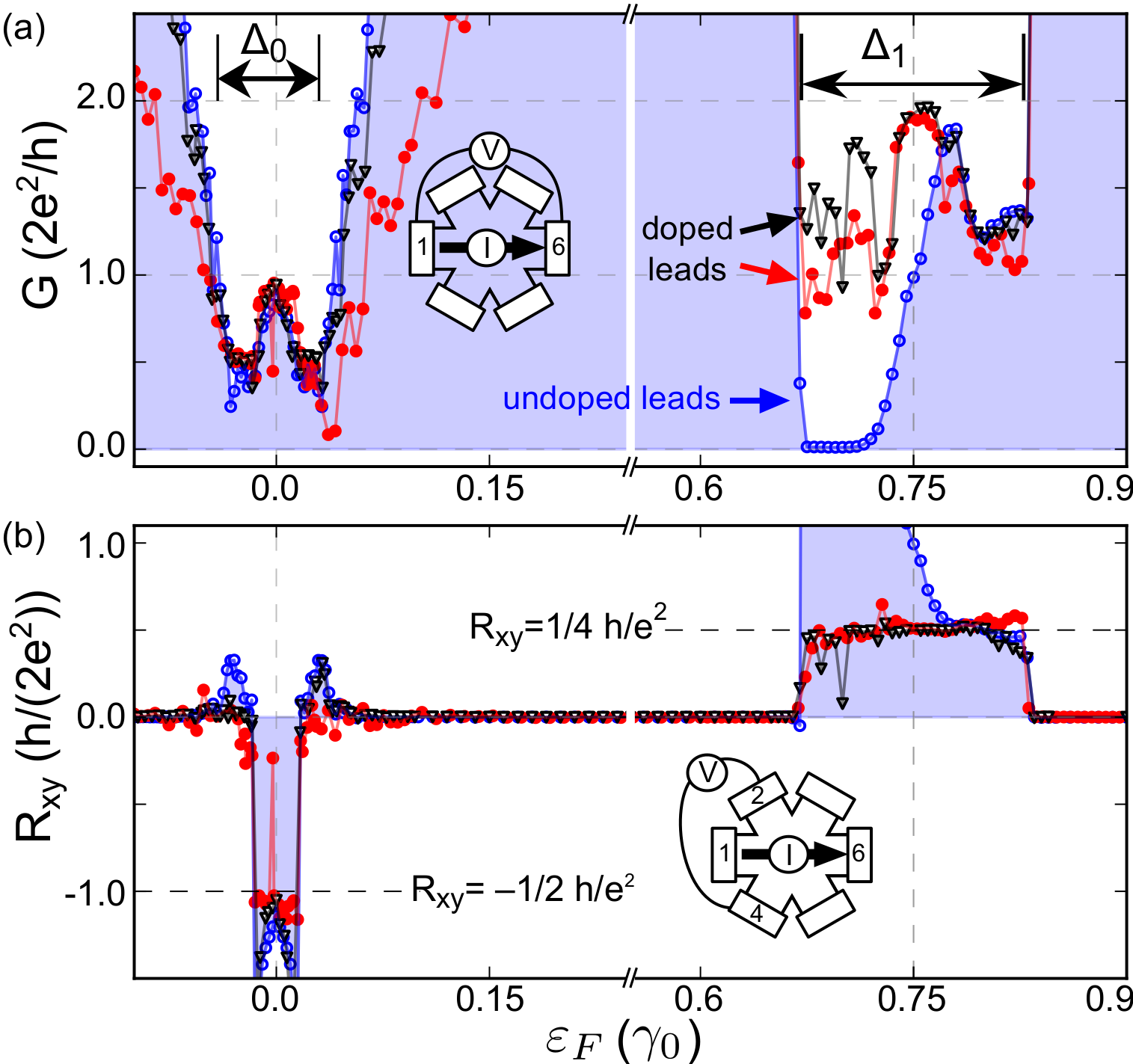}
\caption{(color online) Results for the setup of Fig. \ref{fig1}(c), parameters are as in Fig.\ref{fig2}(b). (a): Conductance, (b): Hall resistance (see insets), empty (blue) circles are for undoped leads, (black) triangles and full circles are for n-doped leads (onsite energies are shifted by 0.75 and 1.25 $\gamma_0$ respectively).
}
\label{fig3}
\end{figure}

\textit{Hall conductance of irradiated graphene.--} Now we turn to the Hall configuration of Fig. \ref{fig1}(c), where the hexagonal symmetry guarantees that there is no pumped current. Fig. \ref{fig3}(a) and (b) show, respectively, the conductance between terminals $1$ and $6$, and the Hall resistance measured between $2$ and $4$ (terminals $1$ and $6$ are the current source and drain, respectively). Empty circles with shaded area underneath is for undoped graphene leads whereas black (triangles) and red (solid circles) traces are for highly doped leads. Besides the strong modulation of the conductance/resistance, the Hall resistance presents three main features: (i) a non-vanishing Hall response at the Floquet gaps, (ii) the Hall resistance for $\varepsilon_F\sim 0$ has the opposite sign than for $\varepsilon_F\sim \hbar\Omega/2$, and (iii) a strong dependence of the modulation on doping. (i) is produced by the Floquet chiral edge states \cite{Oka2009,Perez-Piskunow2014,Usaj2014}, while (ii) follows from the fact that the chiral states have opposite velocities at each gap---changing the chirality of the radiation field's polarization changes the sign of the velocity in both gaps.

Although the contrast between the conductance and Hall resistances just below and above the center of the dynamical gap ($E\sim\hbar\Omega/2$)  (Figs. \ref{fig3}(a) and (b)) may look startling at first sight, this can be understood as the result of a mismatch between states inside and outside the radiated region. Incident electrons coming from the leads belong to a  well defined Floquet  channel (say $n=0$) and then couple with the Floquet states inside the radiated region through their projection on the $n=0$ replica. Mismatch occurs because the states in the undoped lead have a projection of their pseudo-spin along the ribbon which remains roughly constant across the gap while the corresponding pseudospin of the $n=0$ projected part of the Floquet edge state flips sign at the center of the gap, being parallel to that in the non-radiated area for $E\sim\hbar\Omega/2^+$ and antiparallel for $E\sim\hbar\Omega/2^-$ \cite{Usaj2014} \cite{supp-info}. This gives the anomalous conductance suppression with the `S' shape in Fig. \ref{fig3}(a). 
Doping the leads alleviates this mismatch since above the van Hove singularity graphene behaves as a normal metal. This is indeed what we observe in Figs. \ref{fig3}(a) and (b) (black and red lines) when the leads are heavily doped. The Hall resistance reaches roughly constant values. 

\begin{figure}[btp]
\includegraphics[width=.95\columnwidth]{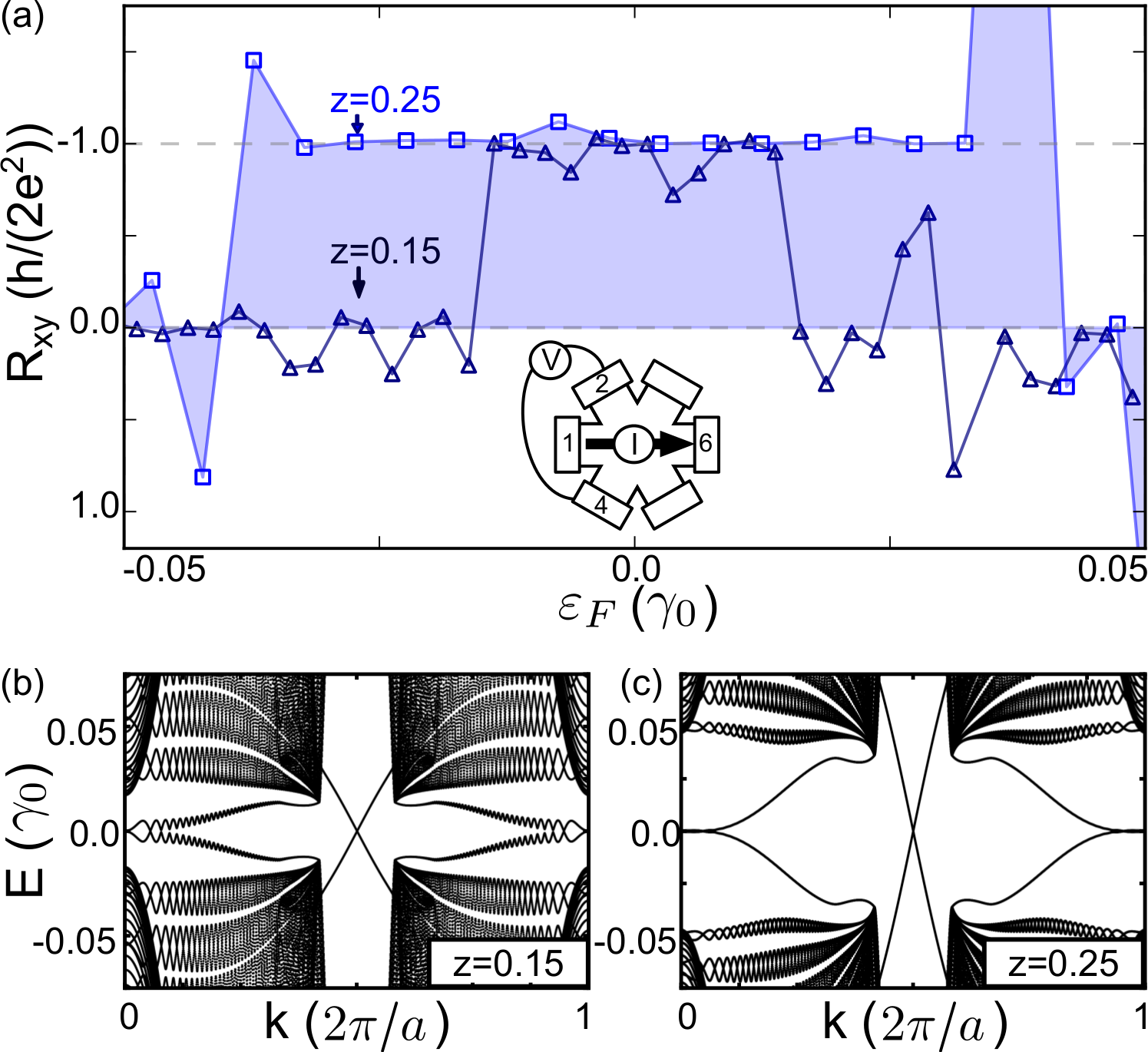}
\caption{(color online) (a) Detail of the Hall resistance in the hexagonal setup close to the Dirac point. Triangles (squares) are for laser intensity parametrized by $z=0.15$ ($z=0.25$). The results are for $\hbar\Omega=1.5 \gamma_0$ and zigzag terminated leads ($W=99a$) with onsite energies are shifted by 1.5 $\gamma_0$). The full quasi-energy spectra for these intensities is shown in (b) and (c). Chiral edge states other than those shown in the projection on the Floquet channel $n=0$ in Fig. 2(a) also develop but they are not impacting the Hall resistance.}
\label{fig4}
\end{figure}

In static systems, Quantum Hall plateaus are determined by the number of chiral edge states which, by means of the bulk-boundary correspondence, can be related to the Chern numbers of the Bloch bands \cite{Hasan2010}. In FTIs, the Chern numbers might not be enough and other topological invariants could be needed to determine if the material is topologically trivial or not \cite{Rudner2013}. But in any case, an explicit calculation (see Fig. \ref{fig4}(b) and (c) shows that there are indeed chiral edge states around $E\sim0$ other than those showing up in Fig.\ref{fig2}(a) but which are not impacting the Hall conductance. Indeed, besides the states crossing at $k=\pi/a$ with zero energy (which are clearly distinguished in the dispersion projected on the $n=0$ channel shown in Fig. \ref{fig2}(a)), in Figs. \ref{fig4}(b) and (c) one can see that other chiral edge states develop. Their projection on $n=0$ is, however, negligible.
Fig.\ref{fig4}(a) shows the dc Hall resistance for small to moderate laser intensities parametrized by $z$. The leads are doped to lessen matching problems. For $z=0.25$ the plateau becomes broader but we observe no change in its value. \textit{This shows that for Floquet topological insulators in a scattering configuration, the chiral edge states do not stand all on the same footing in determining the Hall conductance plateaus}.
 
Rather, our results support that it is only the chiral edge states adding to the dc density of states (the $n=0$ channel) which determine the value of the conductance plateau (one state for $E \sim 0$ and two for $E \sim \pm \hbar\Omega/2$). Otherwise, their contribution vanishes after the time-averaging needed to compute the dc response. Indeed, for the weak driving regime discussed here the chiral edge states not showing up in the dc density of states (encoded in the color scale in Fig. \ref{fig2}(a)) are very poorly coupled to the $n=0$ channel and transport through them is inhibited. Thus, a hierarchy for the Floquet chiral edge states emerges both in the dc density of states \cite{Usaj2014} and in the Hall response, as shown here. This represents a major departure between Floquet topological insulators and systems with time-independent Hamiltonians.

Although the chiral edge states away from $k\sim\pi/a$ do not contribute to the Hall response, the width of the Hall plateaus in Fig. \ref{fig4}(a) might be reduced because of photon-assisted processes involving regions with a large density of states in the higher-order bands. This is not observed at the dynamical gap which is linear in $z$ and therefore better protected from higher-order processes. 
In the opposite limit of very high laser intensities the physics may change completely since the states become highly delocalized along the Floquet channel $n$ and the hierarchy is lost. 
This limit is unrealistic for graphene, but might become available in other materials/systems and is an interesting problem for further study.

\textit{Final Remarks.--} In summary, based on a Floquet scattering picture we addressed the multi-terminal dc conductance of driven systems. Our numerical results for the laser-induced Floquet chiral edge states in graphene show that the Hall resistance reaches plateaus when tuning the interfaces with the leads. The Hall plateaus are found not to follow the usual connection with the number/chirality of the Floquet edge states. Many of these states remain ``silent" and only those states weighting on the dc density of states do contribute.
Whether this hidden edge states could manifest in noise correlations or the time dependent current remains an open and interesting issue.

\clearpage

%\begin{center} 

\section{Supplementary Material}
%\end{center}
\medskip
\smallskip

\setcounter{section}{0}
\setcounter{subsection}{0}
\setcounter{figure}{0}
\setcounter{equation}{0}
\setcounter{NAT@ctr}{0}

\renewcommand{\figurename}[1]{Fig. }
\renewcommand{\theequation}{S\arabic{equation}}

\textbf{Simulation scheme.--} The transport simulations were carried out using Floquet scattering theory \cite{Moskalets2002,Kohler2005} as outlined in the main text. The transmission probabilities in Eqs. (1-3) can be computed from the Floquet-Green's functions as described in Chapter 6 of Ref. \onlinecite{FoaTorres2014}.

Frequencies in the mid-infrared range may offer better chances of experimental realization \cite{Perez-Piskunow2014,Usaj2014} but would require a much higher computational cost because a larger system is needed for the chiral edge states to develop at small to moderate laser intensities. The parameters used here are similar to those in the figures of the main text and are chosen for illustration purposes.

\begin{figure}[b]
\includegraphics[width=\columnwidth]{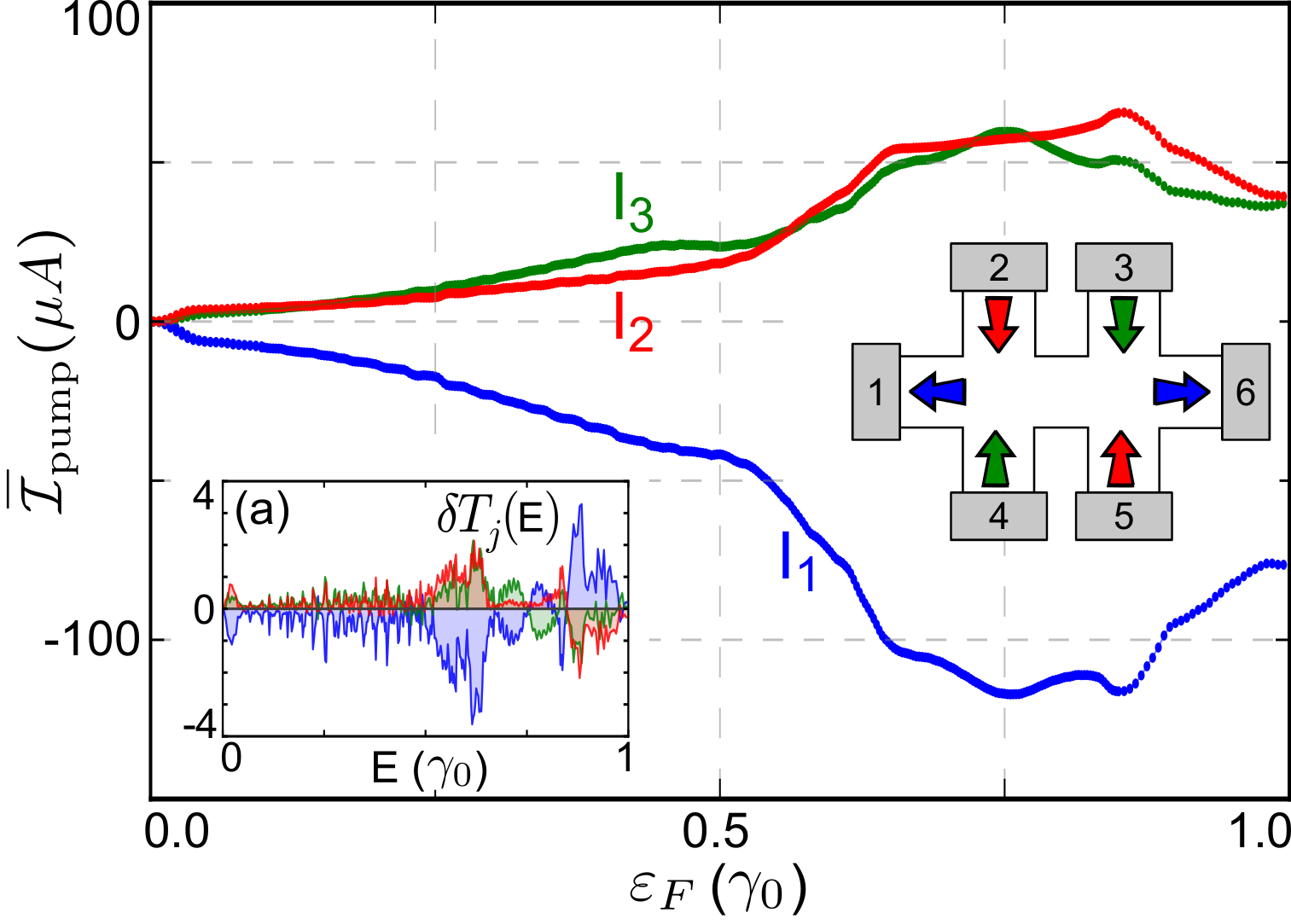}
\caption{
%\textbf{Emergence of laser-induced edge states}. 
(main frame) Results for an H-shaped six terminals configuration as shown in the scheme. (inset a) Directional asymmetry in the transmission coefficients for terminals 1 to 3. These results are for undoped graphene leads of width $W=99a$ (1 and 6 are zigzag ribbons, 2-5 are armchair). The radiated area is a square of length $L=512a$. The laser of frequency $\hbar\Omega=1.5\gamma_0$ is turned-off slowly over a length of $30a$ and the intensity is parametrized by $z=0.15$.}
\label{figHsetup}
\end{figure}

\textbf{Pumped current in an H-shaped six terminal configuration.--} As mentioned in the text, in a multi-terminal setup inversion symmetry alone does not warrant a zero pumped current. This is specially evident in a graphene sample with the six terminal configuration of Fig. 1(a) of the main text. Figure \ref{figHsetup} shows numerical results for the pumped current through each lead in such a setup. We show only the currents in leads 1, 2 and 3 as represented in the scheme. The currents in the remaining leads overlap with those shown: $\overbar{{\cal I}}_6=\overbar{{\cal I}}_1$, $\overbar{{\cal I}}_2=\overbar{{\cal I}}_5$ and $\overbar{{\cal I}}_3=\overbar{{\cal I}}_4$. The inset (a) shows the asymmetry of the corresponding transmission coefficients $\delta {\cal T}_j(\varepsilon) = \sum_i({\cal T}_{i,j}(\varepsilon) - {\cal T}_{j,i}(\varepsilon))$.

\begin{figure}[b]
\includegraphics[width=0.8\columnwidth]{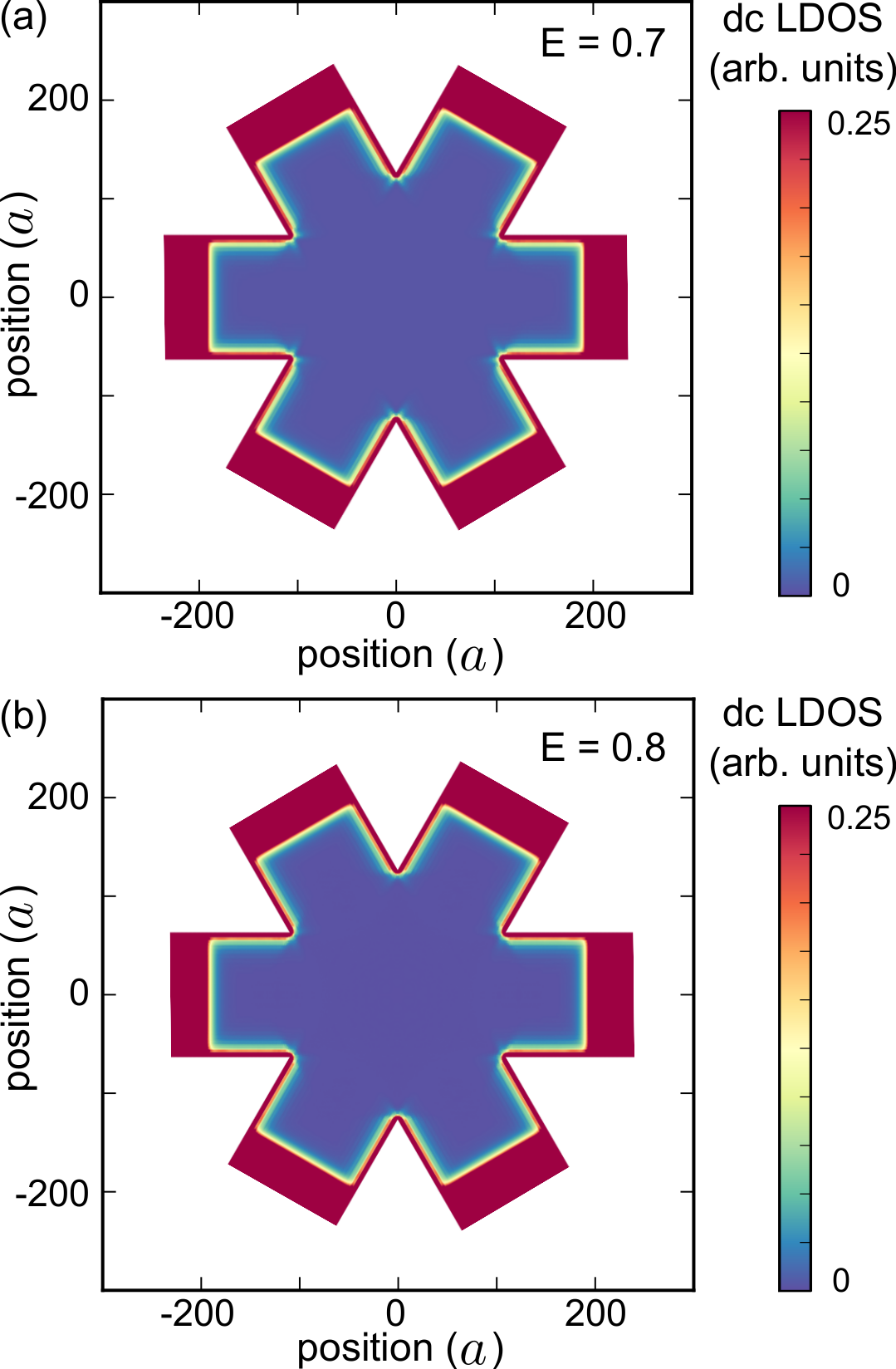}
\caption{
%\textbf{Emergence of laser-induced edge states}. 
Time-averaged local density of states for the hexagonal configuration used in Fig. 3 of the main text at two different energies (the same as in Fig. 3(c) and (d)) (a) $E=0.7\gamma_0$ and (b) $E=0.8\gamma_0$. While delocalized states are available in the leads, the edge states and the bulk gap are clearly observed within the central part (illuminated area). The parameters are the same as in Fig. 3.}
\label{figLDOS}
\end{figure}

\textbf{Local density of states at the dynamical gap.--} Matching problems between the illuminated sample and the undoped and non-illuminated leads were shown in the main text to lead to a decrease in the two-terminal conductance following an `S' shape. This reduction/asymmetry of the electronic transmission is not due to the lack of available states as shown in Fig. \ref{figLDOS}. Fig. \ref{figLDOS} shows the dc local density of states at two selected energies (a) $E=0.7\gamma_0$ and (b) $E=0.8\gamma_0$. In consistency with analytical calculations \cite{Perez-Piskunow2014,Usaj2014} no appreciable differences between the two panels are observed, the contribution from the chiral edge states in the illuminated part of the sample is clear.

\begin{figure}[tbp]
\includegraphics[width=\columnwidth]{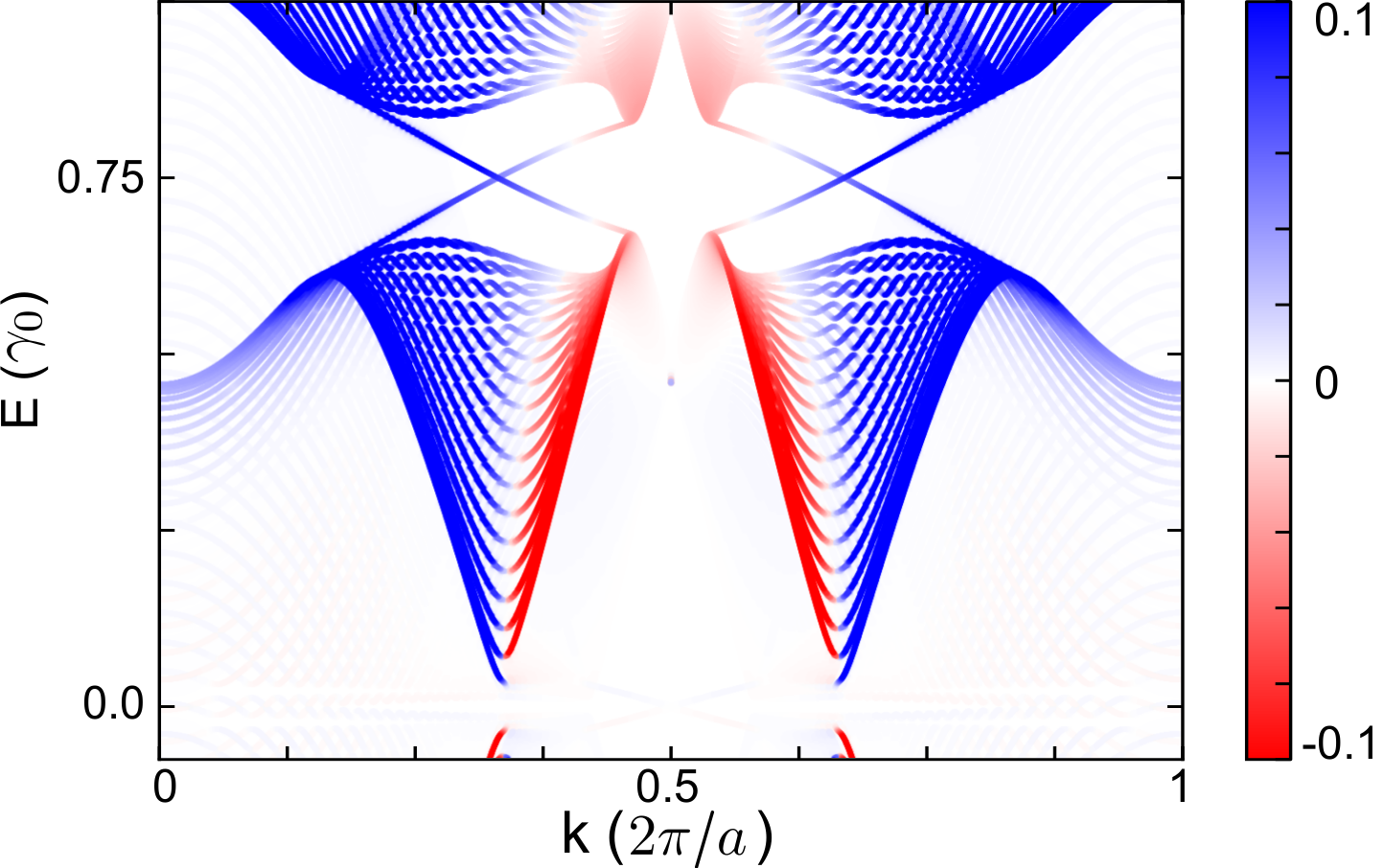}
\caption{(a) and (b): Quasi-energy band structure for a zigzag ribbon of width $W=125a$, in presence of a laser of frequency $\hbar\Omega=1.5 \gamma_0$ and $z=0.1$. The color scale indicates the expectation value of the pseudospin projected along the ribbon direction and the $n=0$ Floquet channel. The color scale is saturated above/below $0.1$ for better  .}
\label{fig-pseudospin}
\end{figure}

\textbf{The role of pseudospin on the matching between non-illuminated (and undoped) and illuminated areas.--} Now we present some results to support our statement that the pseudospin plays a role on the matching problems found in the numerics for the case of undoped graphene leads. Specifically we computed the projection of the pseudospin along the ribbon direction (where translational invariance holds) in the presence of laser illumination, considering only the $n=0$ part of the Floquet eigenfunction. The results are shown in Fig. \ref{fig-pseudospin}, the color scale indicates the value of the pseudospin projection on the corresponding eigenstate. If we sweep the dynamical gap from top to bottom, we can see that pseudospin changes sign when passing from the bottom part of the dynamical gap to the top. Since the pseudospin of electrons incoming from the non-radiated and undoped graphene leads keep their pseudospin projection constant, a mismatch is expected. Transport will then be suppressed on the lower part of the dynamical gap and favoured on the upper part.

\begin{figure}[t]
\vspace{0.5cm}
\includegraphics[width=\columnwidth]{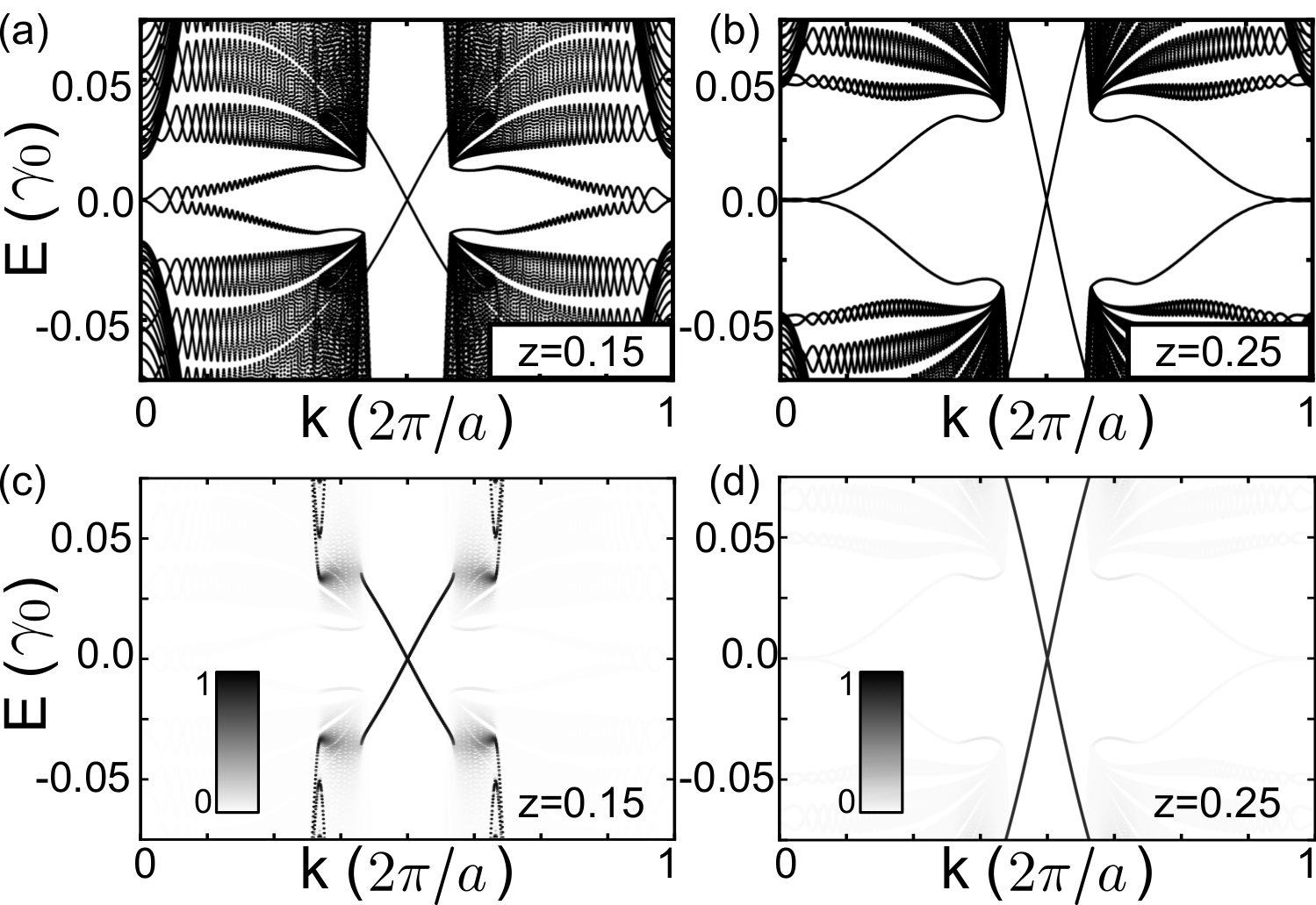}
\caption{(a) and (b): Full quasi-energy band structure for two laser intensities, $z=0.15$ (a) and $z=0.25$ (b). (c) and (d) represent the same data but with a color scale showing the weight of the corresponding eigenstate on the $n=0$ Floquet channel. White is for zero weight and black is for unit weigth. These results are for $\hbar\Omega=1.5 \gamma_0$ and zigzag terminated leads with $W=99a$.}
\label{fig-weightn0}
\end{figure}

\textbf{Detail of chiral edge states developing at higher order replicas and with $E\sim0$.--} In the discussion around Fig. 4 we mentioned that only the chiral edge states crossing at $k=\pi/a$ have an important weight on the $n=0$ channel. In Fig. \ref{fig-weightn0} we show this explicitly. Figs. \ref{fig-weightn0}(a) and (b) reproduce Figs. 4(b) and (c) of the main text and show the full Floquet quasi-energy structure. Figs. \ref{fig-weightn0}(c) and (d) show the same quasi-energy structure in a color scale where the color encodes the weight on the $n=0$ channel, from zero weight (white) to unit weight (black). The chirality of the edge states away from $k=\pi/a$ is the opposite as those crossing at $k=\pi/a$.

\vspace{0.5cm}
\textit{Acknowlegdments.--} We acknowledge financial support from ANPCyT (PICTs 2006-483, 2008-2236, 2011-1552 and
2010-1060), and PIPs 11220080101821 and 11220110100832 from CONICET. LEFFT and GU acknowledge the support of Trieste's ICTP and of the Alexander von Humboldt and Simons Foundations, respectively. Computing time from CCAD-UNC is acknowledged. \\


\begin{thebibliography}{47}
\bibitem [{not()}]{nota1}%
  \BibitemOpen
  \href@noop {} {}\bibinfo {note} {This can be done through a self-consistent
  procedure.}\BibitemShut {Stop}%
\end{thebibliography}
\end{document}